\documentclass[12pt]{article}

\def\lromn#1{\uppercase\expandafter{\romannumeral#1}}

\usepackage{amssymb}
\usepackage{amsmath}
\usepackage{graphicx}
\usepackage{latexsym}

\begin{document}

\begin{titlepage}

\begin{center}

\hfill KEK-TH-1096 \\
\hfill \today

\vspace{1cm}
{\large\bf Solving cosmological problem in universal extra dimension models by
introducing Dirac neutrino}
\vspace{1.5cm}

{\bf Shigeki Matsumoto}$^{a,\,}
$\footnote{smatsu@post.kek.jp},
{\bf Joe Sato}$^{b,\,}
$\footnote{joe@phy.saitama-u.ac.jp},
{\bf Masato Senami}$^{c,\,}
$\footnote{senami@icrr.u-tokyo.ac.jp}, \\
and 
{\bf Masato Yamanaka}$^{b,\,}
$\footnote{masa@krishna.phy.saitama-u.ac.jp} \\

\vskip 0.15in

{\it
$^a${Theory Group, KEK, Oho 1-1, Tsukuba, Ibaraki 305-0801, Japan} \\
$^b${Department of Physics, Saitama University, 
        Shimo-okubo, Sakura-ku, Saitama, 338-8570, Japan} \\
$^c${ICRR, University of Tokyo, Kashiwa 277-8582, Japan }
}

\vskip 1in

\abstract{
Universal extra dimension (UED) models with right-handed neutrinos are studied.
The introduction of the neutrinos makes us possible not
only to describe Dirac neutrino masses but also to solve the cosmological problem
called the KK graviton problem. This problem is essentially caused by the late
time decay of a KK photon into a KK graviton and a photon,
and it distorts the spectrum of the cosmic microwave background or the diffuse photon.
We point out that, once we introduce right-handed neutrinos to UED models,
the KK photon decays dominantly 
into neutrinos and does not emit a photon. We also discuss sub-dominant
modes with a photon in the decay quantitatively,
and show that their branching ratios are so small that
the spectra are not distorted.}

\end{center}
\end{titlepage}
\setcounter{footnote}{0}

\vspace{1.0cm}
\lromn 1 \hspace{0.2cm} {\bf Introduction}
\vspace{0.5cm}

Investigating the nature of dark matter in our universe is important for
particle physics, cosmology and astronomy. Many candidates for dark matter has
been proposed so far in the context of physics beyond the standard model (SM).
Among those, universal extra dimension (UED) models \cite{Appelquist:2000nn}
are one of the attractive scenario which provide a good candidate for
non-baryonic cold dark matter naturally. In this scenario, all particles in the
SM can propagate in those extra dimensions. Furthermore, UED models have
Kaluza-Klein (KK) parity conservation, which is a remnant of the KK number
conservation originated in the momentum conservation along an extra dimension.
Under this parity, particles with odd KK number have odd charge, while the
others have even ones. As a result, the lightest KK particle (LKP) is
stabilized by this parity and is a good candidate for dark matter.

The model most extensively studied in this scenario is the minimal UED (MUED)
model, which is defined in five dimensions. The extra dimension
is compactified on an $S^1 / Z_2$ orbifold for producing the SM chiral fermions.
The mass spectrum of KK particles has been precisely calculated in
Ref.~\cite{Cheng:2002iz} including radiative corrections.
They have found that the LKP is the first KK particle of the photon, which is dominantly
composed of the KK particle of the hypercharge gauge boson. In spite of the
simple framework, this model is phenomenologically successful. In fact, it is
consistent with the recent results of $b\rightarrow s\gamma$ \cite{Agashe:2001xt},
the anomalous muon magnetic moment \cite{Agashe:2001ra,Appelquist:2001jz},
$Z\rightarrow b\bar{b}$ \cite{Appelquist:2000nn,Oliver:2002up}, $B-\bar{B}$
oscillation \cite{Chakraverty:2002qk}, $B$ and $K$ meson decays
\cite{Buras:2002ej}, and electroweak precision measurements 
\cite{Appelquist:2000nn,Appelquist:2002wb} when the compactification scale
$1/R$ is large enough, $1/R \gtrsim$ 400 GeV.

In addition, cosmological aspects such as the thermal relic abundance of the
LKP dark matter has also been studied extensively
\cite{Servant:2002aq,Kakizaki:2006dz,Matsumoto:2005uh}.
The recent work 
\cite{Kakizaki:2006dz} gives the most precise calculation about the abundance
by including all second KK resonance processes in all coannihilation modes.
From the analysis in the work, it is shown that the compactification scale,
which is consistent with the WMAP observation \cite{WMAP}, is 600 GeV
$\lesssim 1/R \lesssim$ 1400 GeV. Furthermore, the upper bound of the Higgs
mass has been obtained as $m_h \lesssim$ 230 GeV.

Although phenomenologically successful, this model has two shortcomings.
First one is the absence of neutrino masses. As well known, neutrino oscillation
experiments indicate the existence of neutrino masses \cite{Fogli:2005cq}.
However, neutrinos are massless in the MUED model,
because this model is the simple extension of the SM to the extra dimension.
Hence, the model is required to extend to describe neutrino masses.
Second issue concerns the cosmological aspect of this model.
For $1/R \lesssim$ 800 GeV, it has been pointed out that
the KK particle of the graviton is the LKP \cite{Kakizaki:2006dz,Matsumoto:2005uh}.
This is a serious problem for the model,
because the lifetime of the next LKP (NLKP), that is the KK photon,
is very long, and its decay in the early universe leads to an inconsistency with the
observation of the cosmic microwave background (CMB)
\cite{Hu:1993gc,Feng:2003xh} or the diffuse photon \cite{Feng:2003xh}.

In this letter, we show that these two problems can be solved simultaneously
by introducing right-handed neutrinos.
As we will discuss in section \lromn 3, tiny Yukawa couplings are
necessarily introduced to describe neutrino masses in the framework of UED models.
Then the first KK particles of right-handed neutrinos are the NLKPs
in the region where the KK graviton is the LKP,
because the radiative corrections to the KK right-handed neutrinos
are negligibly small due to the tiny Yukawa couplings.
Therefore, a KK photon decays dominantly into a KK right-handed neutrino
and an ordinary neutrino, and photons are not emitted in the leading process of its decay.
Thus, the CMB and diffuse photon spectra are not distorted.

We discuss the above solution quantitatively.
First, we briefly summarize the KK graviton problem in the next section.
Next, the right-handed neutrinos are introduced to provide
neutrino masses, and some decay widths of the KK photon into
the first KK right handed neutrino are calculated in Sec.\lromn 3.
Then, we discuss that the decay of the
first KK photon does not affect spectra of the CMB and diffuse photon,
and also does not spoil the successful prediction of the big bang
nucleosynthesis (BBN) in Sec.\lromn 4.
Finally, Sec.\lromn 5 is devoted to summary and discussion.

\vspace{1.0cm}
\lromn 2 \hspace{0.2cm} {\bf KK graviton problem in the MUED model}
\vspace{0.5cm}

First, we consider the mass spectrum of first KK particles, which is of
importance for the KK graviton problem. At tree level, the mass of a KK particle
is determined by the compactification scale $1/R$ and the corresponding SM
particle one as $m^{(n)} = (n^2/R^2 + m_{\rm SM}^2)^{1/2}$. Since the
compactification scale is much larger than the SM particle masses, all KK
particles at each KK mode are highly degenerated in mass around $n/R$.
Mass differences among KK particles at each KK mode
dominantly come from radiative corrections,
which are calculated in Ref.~\cite{Cheng:2002iz}.
Thus, for instance, colored KK particles are heavier than other KK particles,
while the masses of KK U(1) gauge boson and right-handed leptons
still remain $\sim n/R$.

The candidates for the LKP are the KK photon ($\gamma^{(1)}$), the KK
charged Higgs ($H^{\pm(1)}$), and the KK graviton ($G^{(1)}$)
depending on $1/R$ and the Higgs mass ($m_h$).
Here $H^{\pm(1)}$ is the KK particle of the Goldstone boson.
Since KK gauge bosons acquire mass term from the fifth components in gauge fields,
KK particles of Goldstone bosons remain as physical states.

\begin{figure}[t]
\begin{center}
\scalebox{1.0}{\includegraphics*{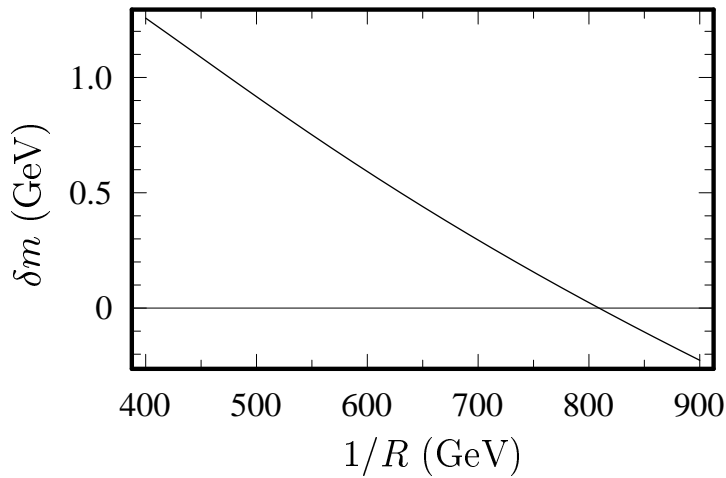}}
\caption{\small The difference, $\delta m \equiv m_{\gamma^{(1)}} - 1/R$, is
shown as a function of $ 1/R$. The cutoff scale $\Lambda$ is set to be
$\Lambda R = 20$.}
\label{fig:deltam}
\end{center}
\end{figure}

First, the mass of the KK photon is
obtained by diagonalizing following mass squared matrix,
\begin{eqnarray}
  \begin{pmatrix}
    1/R^2 + \delta m_{B^{(1)}}^2 + g^{\prime 2}v^2/4
    &
    g'gv^2/4
    \\
    g'gv^2/4
    &
    1/R^2 + \delta m_{W^{(1)}}^2 + g^2v^2/4
  \end{pmatrix}~,
\end{eqnarray}
which is written in the ($B^{(1)}$, $W_3^{(1)}$) basis. Here, $g(g')$ is
SU(2)$_L$ (U(1)$_Y$) gauge coupling constant. Radiative corrections,
$\delta m_{B^{(1)}}^2$ and $\delta m_{W^{(1)}}^2$, are given by
\begin{eqnarray}
  \delta m_{B^{(1)}}^2
  &=&
  -
  \frac{39}{2}
  \frac{g^{\prime 2}\zeta(3)}{16\pi^4R^2}
  -
  \frac{1}{6}
  \frac{g^{\prime 2}}{16\pi^2R^2}\ln(\Lambda^2R^2)~,
  \nonumber \\
  \delta m_{W^{(1)}}^2
  &=&
  -
  \frac{5}{2}
  \frac{g^2\zeta(3)}{16\pi^4R^2}
  +
  \frac{15}{2}
  \frac{g^2}{16\pi^2R^2}\ln(\Lambda^2R^2)~,
\end{eqnarray}
where $\Lambda$ is the cutoff scale of the MUED model, it is usually taken to
be $\Lambda R \sim {\cal O}(10)$. In this work, we adopt the value, $\Lambda R
= 20$. Even if different value of this cutoff scale was adopted, our
conclusion is not changed significantly.
The mass of the KK photon is smaller than $1/R$ for $1/R\gtrsim $ 800 GeV. The
difference, $\delta m \equiv m_{\gamma^{(1)}} - 1/ R $, is shown as a function of
$ 1 / R$ in Fig.~\ref{fig:deltam}. As seen in this figure, $\delta m $ is
typically on the order of 1 GeV and small for larger $1/R$.

Next, the mass of the KK charged Higgs is given as
\begin{eqnarray}
 m_{H^{\pm (1)}}^2
 &=&
 1/R^2 + m_W^2 + \delta m_{H^{(1)}}^2~,
 \nonumber \\
 \delta m_{H^{(1)}}^2
 &=&
 \left(
  \frac{3}{2}g^2 + \frac{3}{4}g^{\prime 2} - \lambda_h
 \right)
 \frac{1}{R^2}
 \frac{\ln\left(\Lambda^2R^2\right)}{16\pi^2}~,
\end{eqnarray}
where $\lambda_h$ is the Higgs self-coupling defined by $\lambda_h \equiv
m_h^2/v^2$. As $m_h$ is increased, the
negative contribution in the radiative correction, $\delta m_{H^{(1)}}^2$,
increases. Hence, for large $\lambda_h$, the mass difference between the KK
photon and $H^{\pm(1)}$ can be negative.

\begin{figure}[t]
\begin{center}
\scalebox{1.0}{\includegraphics*{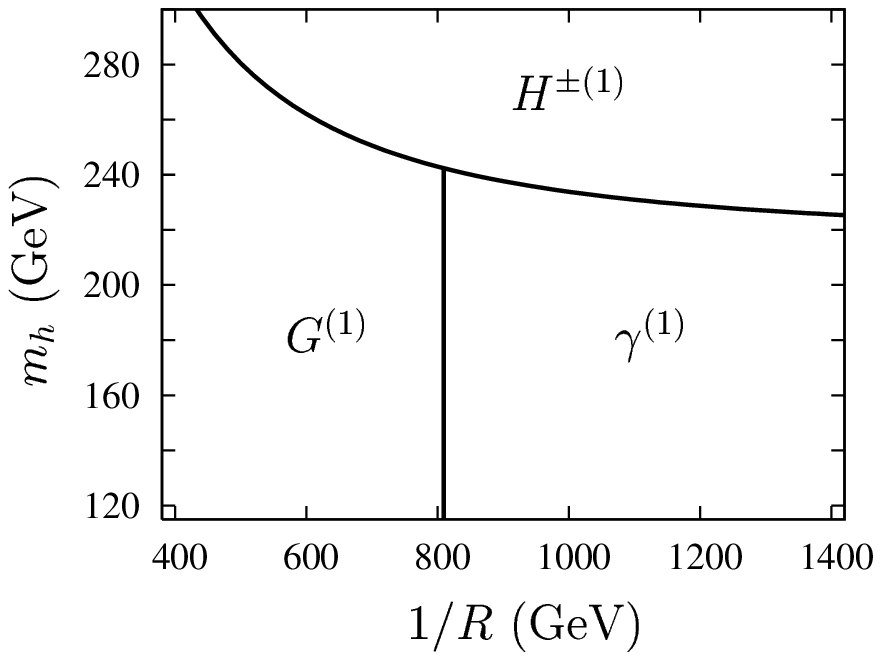}}
\caption{\small The constituent of the LKP in the MUED model is shown.
The cutoff scale $\Lambda$ is set to be $\Lambda R = 20$.}
\label{fig:LKP}
\end{center}
\end{figure}

Finally, we consider the mass of the KK graviton.
Since all interactions relevant to the KK graviton are suppressed by the Planck scale,
radiative corrections are negligible.
Therefore its mass is simply given by $1/R$.
By the comparison with the mass of the KK photon or KK charged Higgs,
it turns out that the KK graviton is the LKP when $1/R$ and
$m_h$ are not so large. We summarize the result in Fig.~\ref{fig:LKP}.
The parameter space in the MUED model is divided into three regions,
where the LKP is $\gamma^{(1)}$, $H^{\pm (1)}$, and $G^{(1)}$, respectively.
Obviously, the $H^{\pm (1)}$ LKP region
is ruled out from the viewpoint of dark matter.
Furthermore, the triviality bound on the Higgs mass term gives the
severer constraint on the charged LKP region. For $\Lambda R = 20$, the
bound requires $m_h \lesssim 200$ GeV, thus the region is completely
disfavored \cite{Bhattacharyya:2006ym}.
For lower cutoff, the bound allows $m_h$ to be large enough,
and the theoretical upper bound on $m_h$ comes from the charged LKP bound.

Moreover, the $G^{(1)}$ LKP region is also excluded by considering cosmological
implication from NLKP decay. When the temperature of the universe is
around $1/R$, the KK photon, which is the NLKP in this region, is in thermal
equilibrium. On the contrary, the KK graviton is not in thermal equilibrium
and its abundance is tiny at that time due to extremely weak
interactions\footnote{We assume that the reheating temperature of the universe
after inflation is so low that the KK graviton is not produced abundantly from
the thermal bath.}. After the temperature of the universe becomes low enough,
the annihilation process of the KK photon is frozen out, and its density per co-moving
volume is fixed. The KK photon decays into the KK graviton long after the
freeze-out through the process, $\gamma^{(1)} \rightarrow G^{(1)} + \gamma$,
which occurs typically in or after the recombination era as shown in the next section.
Since the observation of the CMB spectrum is very accurate and any deviation
from the black-body spectrum has not been observed,
this decay in the recombination era is severely constrained\footnote{
In fact, the constraint is more stringent than
that from the BBN for the decay in the recombination era \cite{Feng:2003xh}.}.
Furthermore, even if the decay takes place after the recombination era,
it produces observable peak in the diffuse photon spectrum.
As a result, the KK graviton decay is severely constrained
from these measurements, and in fact, the entire $G^{(1)}$ LKP region is ruled out.

Fortunately, in the $\gamma^{(1)}$ in Fig.~2 LKP region, where the NLKP is $G^{(1)}$,
there is no KK graviton problem
when the reheating temperature is low enough \cite{Feng:2003nr}.
However, most of the parameter region consistent with the dark matter abundance
observed by WMAP is in the $G^{(1)}$ LKP region \cite{Kakizaki:2006dz}.
Furthermore, this region is more attractive than others for physics at
colliders such as the LHC.

The KK graviton problem is not the intrinsic one in the MUED model. It will
appear in various UED extensions of the SM. Thus, it is important to consider a
mechanism in order to avoid the problem. The mechanism we present in this
letter is very simple and can be applied to other UED models.

Here, it is worth notifying another cosmological problem related to
the radion field. The radion field comes from the higher dimensional
component of the metric and behaves as a scalar field in the four
dimensional view point after the compactification.
The zero mode of the radion field tends to overclose the universe easily,
if the potential along the radion direction is flat enough \cite{Kolb:2003mm}.
This problem is the common problem in extra dimension scenarios,
and occurs not only UED models but also other models extended to TeV
scale higher dimensions. The problem depends on the inflation model,
and it can be solved if the radion mass is large enough \cite{Mazumdar:2003vg}.
Solving the problem is beyond the scope of this letter, and we only
assume that an appropriate inflation model solves the radion problem.

\vspace{1.0cm}
\lromn 3 \hspace{0.2cm} {\bf Introduction of right-handed neutrinos to the MUED
model}
\vspace{0.5cm}

In addition to the KK graviton problem, the MUED model has another problem,
which is the absence of neutrino masses. Since the MUED model is the simple
extension of the SM to the higher dimensions, neutrinos are treated as
massless particles. On the other hand, the existence of neutrino masses and
lepton flavor mixing have been established by neutrino oscillation experiments.
Thus, we have to extend the MUED model in order to construct a more realistic one.

In the framework of UED models,
the neutrino masses are described by Dirac and Majorana mass terms as
\begin{eqnarray}
 {\cal L}_{\nu{\rm -mass}}
 = y_\nu \bar N L \phi
 + M  N^T C \gamma_5 N + \rm h.c. ,
\end{eqnarray}
where $N$ is the right-handed neutrino and
$C$ is the charge conjugation operator, $ C \equiv i \gamma_0 \gamma_2$.
After the compactification,
the second term gives the Majorana mass term to the right-handed neutrino \cite{Pilaftsis:1999jk}.
According to the algebraic structure of the five dimensional space-time,
ordinary Majorana mass terms in the five dimensional space-time, $N^T C N$
are not allowed \cite{Weinberg:1984vb,Pilaftsis:1999jk}.
 
Since the UED model describes physics below the cutoff scale $\Lambda \sim 10$ TeV,
the Majorana mass $M$ should be less than the scale.
As a result, tiny Yukawa couplings are inevitably introduced to explain neutrino masses.
Thus, the seesaw mechanism \cite{seesaw} by these Majorana masses
can not be excellent solution for the smallness of neutrino masses in the UED setup.
In the following, we will consider only Dirac mass terms
and ignore the Majorana mass terms for definiteness\footnote{For example,
if we impose the lepton number symmetry to the model, the absence of Majorana masses is realized.}.
Even if we introduce the Majorana mass terms, our discussion is not changed
as long as the Yukawa coupling is small enough and KK right-handed neutrinos
are out of equilibrium at the early universe. Due to the existence of these tiny couplings,
the KK graviton problem is resolved as discussed in the next section.

Once we introduce right-handed neutrinos in the MUED model,
their KK particles automatically appear in the spectrum.
Since these particles are the SM gauge singlet
and have only small Yukawa interactions,
$y_\nu \sim m_\nu / v$, where $m_\nu$ is
a neutrino mass and $v \sim 246$ GeV is the vacuum expectation value of the
Higgs field,
radiative corrections to their masses are very small.
The mass of the KK particle of a right-handed neutrino,
$N^{(1)}$, is estimated as
\begin{eqnarray}
 m_{N^{(1)}}  \simeq \frac{1}{R} + O \left( \frac{m_{\nu }^2}{1/R} \right)~.
\end{eqnarray}

The KK right handed neutrino is the NLKP in the $G^{(1)}$ LKP region, since the mass
is smaller than that of the KK photon.
The existence of the $N^{(1)}$ NLKP changes the
late time decay of the KK photon. In this section, we consider several decay
modes of the KK photon, which are important for the detailed study of the
thermal history of the universe.

\begin{figure}[t]
\begin{center}
\scalebox{1.0}{\includegraphics*{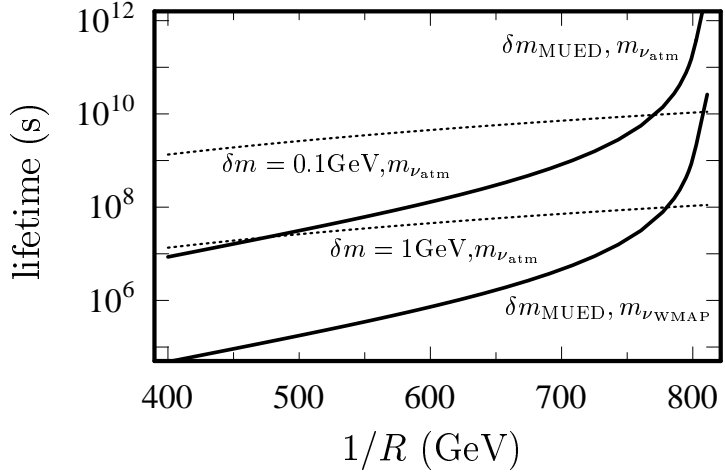}}
\caption{\small The lifetime of $\gamma^{(1)}$ is shown as a function of $1/R$.
In two solid lines, the mass of $\gamma^{(1)}$ is obtained by the MUED model.
The upper line is for $m_\nu = m_{\nu_{\rm atm}}$,
while the lower one is for $m_{\nu_{\rm WMAP}}$.
In dotted lines, $\delta m $ is regarded as a free parameter
and fixing it on 1 (0.1) GeV for the lower(upper) line.
The neutrino mass $m_\nu $ is set to be $ m_{\nu_{\rm atm}}$.}
\label{fig:lifetime}
\end{center}
\end{figure}

The KK photon decays dominantly into $N^{(1)}$ and left-handed neutrino
at tree level. This decay takes place through the diagram in which the KK
left-handed neutrino is flipped into $N^{(1)}$, and its decay width is given by
\begin{eqnarray}
 \Gamma_{N^{(1)} \bar \nu}
 =
 \frac{\alpha'}{8} \frac{m_\nu^2 \delta m^2}{(1/R)^3}
 \simeq
 2\times 10^{-7} ~ [{\rm sec}^{-1}]
 \left(\frac{500 {\rm GeV}}{1/R}\right)^3
 \left(\frac{m_\nu}{0.1 {\rm eV}}\right)^2
 \left(\frac{\delta m}{1 {\rm GeV}}\right)^2 ,
\end{eqnarray}
where $\alpha' = g^{\prime 2}/4\pi$ and the mass difference between
$\gamma^{(1)}$ and $N^{(1)}$ is denoted by $\delta m$. We use the approximation
$m_{N^{(1)}} = m_{G^{(1)}} = 1/R$ unless the mass difference appears. Here,
$m_\nu$ is the heaviest neutrino mass. The lower limit of $m_\nu$ is obtained by
atmospheric neutrino experiments, $m_{\nu_{\rm atm}} \simeq 0.05$~eV
\cite{Fogli:2005cq,SK}
.
We adopt the upper bound from the WMAP observation,
$\sum m_\nu < 2.0$ eV \cite{WMAP,Ichikawa:2004zi}.
In Fig.~\ref{fig:lifetime}, the lifetime of $\gamma^{(1)}$ is shown.
In two solid lines, the mass of $\gamma^{(1)}$ is obtained by the MUED model.
The upper (lower) solid line is the result for
$m_\nu = m_{\nu_{\rm atm}} (m_{\nu_{\rm WMAP}})$,
where $m_{\nu_{\rm WMAP}} = 0.67 $ eV.
Dotted lines are depicted by regarding $\delta m$
as a free parameter and fixing it on 1 (0.1) GeV for the lower(upper) line.
The neutrino mass $m_\nu $ is set to be $ m_{\nu_{\rm atm}}$.

The KK photon can also decay into $G^{(1)}$ through the Planck scale
suppressed interaction. This process may be important for cosmological
implication, because a photon is produced in the
final state. Its decay width is given in Ref.~\cite{Feng:2003xh} as
\begin{eqnarray}
 \Gamma_{G^{(1)}\gamma}
 =
 \frac{5\cos^2\theta_W\delta m^{3}}{9\pi M_{\rm Pl}^2}
 \simeq
 7 \times 10^{-14} ~ [{\rm sec}^{-1}]
 \left(\frac{\delta m}{1 {\rm GeV}}\right)^3 ,
\end{eqnarray}
where $M_{\rm Pl} \simeq 2.4 \times 10^{18}$ GeV is the reduced Planck mass.

In addition to these decay processes, there are two other modes in which a
photon is emitted in the final state. First one is the $\gamma$ associated
$\gamma^{(1)}$ decay into $N^{(1)}$ ($\gamma^{(1)}\rightarrow
N^{(1)}\bar{\nu}\gamma$), which occurs through one-loop diagrams.
Its width is estimated as
\begin{eqnarray}
 \Gamma_{N^{(1)}\bar{\nu}\gamma}
 & \sim &
 \frac{ \alpha^3 \alpha' \sin^2 \theta_W }{ 8 \pi^3 }
 \frac{m_\nu ^2 \delta m^4}{m_{\gamma^{(1)}} m_W^4 } \times 10^{-4}
 \nonumber \\
 &\sim&
 10^{-20} ~ [\text{sec}^{-1}]
 \left( \frac{500 \text{GeV}}{1 / R} \right)
 \left( \frac{m_\nu }{0.1 \text{eV}} \right)^2
 \left( \frac{\delta m}{1\text{GeV}} \right)^4~,
\end{eqnarray}
where $m_W$ is the W boson mass and the factor $10^{-4}$ arises from
loop integrals and phase space.
Thus, this
process is suppressed by one-loop integrals, tiny neutrino mass, and small
phase space. As a result, its width turns out to be much smaller
than the two bodies decay, $\gamma^{(1)}\rightarrow G^{(1)}\gamma$, and we can
neglect it even for the maximum $m_\nu$. Another mode is the decay process
associated with a neutral pion, $\gamma^{(1)}\rightarrow N^{(1)}\bar{\nu}\pi^0$,
in which photons are emitted in the final state through
$\pi^0\rightarrow 2\gamma$ decay. However, this process is also suppressed by
$f_{\pi}^2/m_Z^2$, neutrino mass, and three bodies phase space. Hence, this
decay width is much smaller than the two bodies decay, and we can also neglect
the process.

After all, the decay of the KK photon is governed by the process,
$\gamma^{(1)}\rightarrow N^{(1)}\bar{\nu}$. On the other hand, the dominant
decay mode associated with a photon comes from the Planck suppressed process,
$\gamma^{(1)}\rightarrow G^{(1)}\gamma$, and its branching ratio is
\begin{eqnarray}
 {\rm Br_{X\gamma}}
 =
 \frac{\Gamma_{G^{(1)}\gamma} }{\Gamma_{N^{(1)} \bar \nu}}
 =
 5 \times 10^{-7}
 \left(\frac{1/R}{500 {\rm GeV}}\right)^3
 \left(\frac{0.1 {\rm eV}}{m_\nu}\right)^2
 \left(\frac{\delta m}{1 {\rm GeV}}\right) ~.
\label{eq:br}
\end{eqnarray}
Hence, the decay associated with a photon is very suppressed.

\vspace{1.0cm}
\lromn 4 \hspace{0.2cm} {\bf $N^{(1)}$ dark matter and the KK graviton problem}
\vspace{0.5cm}

In the early universe ($T \sim 1/R$), $\gamma^{(1)}$ is in thermal equilibrium,
while $N^{(1)}$ and $G^{(1)}$ are out of equilibrium due to the tiny couplings.
Thus, their abundances are negligibly small at that time, and they can be
produced only by $\gamma^{(1)}$ decay.
The KK photon decays dominantly into $N^{(1)}$
at $T\sim (\Gamma_{N^{(1)}\bar\nu} M_{\rm Pl})^{1/2}$.
However, $N^{(1)}$ can not 
decay into $G^{(1)}$, since it is kinematically forbidden. As a result, $N^{(1)}$
remains as a non-baryonic cold dark matter in the present universe. Since the
mass difference between $N^{(1)}$ and $\gamma^{(1)}$ is negligibly small, the
allowed region consistent with the WMAP observation is the same as that of the
$\gamma^{(1)}$ LKP.

As discussed in the previous section, $\gamma^{(1)}$ can decay into the final state
with a photon, though its branching ratio is small. Since constraints from
observation of background photons to the late time decay is very stringent, we
should quantitatively check that the decay does not really affect the photon
spectrum. In the following, we consider cosmological implications of the MUED
model with the right-handed neutrinos.
The decay of $ \gamma^{(1)}$ occurs at $t =
10^5 - 10^{12}$~s as shown in Fig.~\ref{fig:lifetime},
it corresponds to the period between the BBN and the recombination.
As a result, the
constraint from measurements of the diffuse photon spectrum can be neglected.
On the other hand, emitted photons may spoil the successful prediction of the
BBN by destroying light nuclei.
Moreover the photons may distort the spectrum of the CMB.
Thus, it is important to consider constraints from measurements of
the BBN and CMB to this model.

The distortion of the CMB spectrum is
parametrized by chemical potential $\mu$
when a process changing the energy of a background photon are effective,
or the Compton $y$-parameter after the process become ineffective.
These values are constrained as
$|\mu| < 9 \times 10^{-5}$ and $|y| <1.2 \times 10^{-5}$ \cite{muy}.
The total injection energy from the decay is used as the quantity constrained
from the CMB measurement. In fact, $\mu$ and $y$ are proportional to the energy.
Furthermore, the constraint from the BBN is also represented using the energy.
Since the energy of a emitted photon is rapidly redistributed through
inverse Compton scattering ($\gamma e^-\to\gamma e^-$) and $e^+e^-$ creation
with background photons ($\gamma\gamma_{\rm BG}\to e^+e^-$),
the important quantity is the total injection energy and not its spectrum.

The injection energy is given by $\epsilon {\rm Br}_{X\gamma} Y_{\gamma^{(1)}}$,
where $\epsilon$ is the typical energy of emitted photon and
$Y_{\gamma^{(1)}}=  Y_{N^{(1)}} = n_{N^{(1)}} / n_{\gamma_{\rm BG}}$
is the number density of $\gamma ^{(1)}$ normalized by
that of background photons.
The branching ratio is given in Eq.~(\ref{eq:br}) and 
$\epsilon$ is less than $\delta m$.
The yield $Y_{N^{(1)}}$ is
estimated by requiring that the relic abundance of $N^{(1)}$ accounts for the
observed abundance of dark matter,
\begin{eqnarray}
 Y_{N^{(1)}}
 =
 \frac{\Omega_{\rm DM} \rho_c}{1/R}
 \frac{1}{n_0} 
 \simeq
 5 \times 10^{-12}
 \left(\frac{\Omega_{\rm DM}h^2}{0.10}\right)
 \left(\frac{500 \rm GeV}{1/R}\right)~,
\end{eqnarray}
where $\rho_c = 1.1\times 10^{-5} h^2~{\rm GeVcm}^{-3}$ is the critical density
of the universe, and $n_0 = 410 ~{\rm cm}^{-3}$ is the number density of
background photons in the present universe. Therefore, the total injection
energy is estimated as
\begin{eqnarray}
 \epsilon {\rm Br}_{X\gamma} Y_{N^{(1)}}
 \lesssim
 3\times 10^{-18} ~ {\rm GeV}
 \left(\frac{1/R}{500 {\rm GeV}}\right)^2
 \left(\frac{0.1 {\rm eV}}{m_\nu}\right)^2
 \left(\frac{\delta m}{1 {\rm GeV}}\right)^2
 \left( \frac{\Omega_{\rm DM} h^2 }{0.10} \right) .
\end{eqnarray}

The successful BBN and CMB scenarios are not disturbed
unless this value exceeds $10^{-9} - 10^{-13}$ GeV \cite{Feng:2003xh}.
The prediction of the MUED model with right-handed neutrinos
is several order of magnitude smaller than the bound.
Therefore, the KK graviton problem is avoided in the MUED model
with right-handed neutrinos.
Moreover, this results is expected to hold for
other UED models with right-handed neutrinos.

\vspace{1.cm}
\lromn 5 \hspace{0.2cm} {\bf Summary and discussion}
\vspace{0.5cm}

In this letter, we have pointed out that the introduction of right-handed
neutrinos resolves two shortcomings of the MUED model simultaneously,
which are the absence of neutrino masses and the KK graviton problem.
A KK photon
decays dominantly into a KK right-handed neutrino and an ordinary neutrino.
In other words, no photon is emitted by the KK photon decay,
and hence this model are free from constraints on the late time decay from the BBN, CMB,
and diffuse photon measurements.
With the calculation of the relic abundance of the KK photon,
this fact allows us to
consider a small compactification scale in the MUED model ($1/R\sim$ 600 GeV),
which is consistent with all results in particle physics experiments
and observed abundance of dark matter.

The dark matter in this region is the KK right-handed neutrino and difficult to
observe a signal in detection measurements for dark matter.
However, smaller value of $1/R$ has a great advantage for collider experiments.
For instance, not only first KK particles but also second KK ones
can be produced at the LHC,
and we can easily find signals of second KK gauge bosons
using energetic dilepton channels \cite{Datta:2005zs}.
Furthermore, the MUED model has only a few undetermined parameters,
and these values 
will be observed in a good accuracy even at the LHC. Once the values are
determined, we can predict the relic abundance of dark matter theoretically,
and can discuss the connection between collider physics and cosmology by
comparing the prediction with WMAP and Planck results \cite{unknown:2006uk}.
The comparison will be a good tool to make sure that the UED model
provides dark matter in our universe.

\vspace{1.0cm}
\hspace{0.2cm} {\bf Acknowledgments}
\vspace{0.5cm}

The work of SM, JS and MS are supported in part by the Grant-in-Aid for the Ministry of
Education, Culture, Sports, Science, and Technology, Government of Japan
(No. 16081211. for SM, No. 17740131 and 18034001 for JS and No. 18840011 for MS).

\end{document}